# Internet of Things Security and Forensics: Challenges and Opportunities

Mauro Conti[1], Ali Dehghantanha[2], Katrin Franke[3], Steve Watson[4]*

1- Department of Mathematics, University of Padua, Italy
2- Department of Computer Science, University of Salford, UK
3- Testimon Forensics Group, Norwegian University of Science and Technology, Norway
4- VTO Labs, Denver, Colorado, USA

**Abstract**
The Internet of Things (IoT) envisions pervasive, connected, and smart nodes interacting autonomously while offering all sorts of services. Wide distribution, openness and relatively high processing power of IoT objects made them an ideal target for cyber attacks. Moreover, as many of IoT nodes are collecting and processing private information, they are becoming a goldmine of data for malicious actors. Therefore, security and specifically the ability to detect compromised nodes, together with collecting and preserving evidences of an attack or malicious activities emerge as a priority in successful deployment of IoT networks. In this paper, we first introduce existing major security and forensics challenges within IoT domain and then briefly discuss about papers published in this special issue targeting identified challenges.

**Keywords:**
Internet of Things, Cyber Security, Digital Forensics

## 1. Introduction

The Internet of Things (IoT) integrates various sensors, objects and smart nodes that are capable of communicating with each other without human intervention [1] . The objects/things function autonomously in connection with other objects. IoT nodes are capable of delivering lightweight data, accessing and authorizing cloud-based resources for collecting and extracting data and making decisions by analysing collected data. The emergence of IoT has led to pervasive connection of people, services, sensors and objects. IoT devices are now deployed in a wide range of applications from smart grids to healthcare and intelligence transport systems [2]. Huge business opportunities that exist within IoT domain significantly increased number of smart devices and intelligent, autonomous services offered in IoT networks. Moreover, reliance of IoT devices on cloud infrastructure for data transfer, storage and analysis led to development of cloud-enabled IoT networks [3].

Security issues such as privacy, access control, secure communication and secure storage of data are becoming significant challenges in IoT environment [4]. Moreover, every single device that we create, every new sensor that we deploy, and every single byte that is synchronized within an IoT environment may at some point come under scrutiny in the course of an investigation.

The fast growth of IoT devices and services led to deployment of many vulnerable and insecure nodes [5]. Moreover, conventional user-driven security architectures are of little use in object-driven IoT networks [6]. Therefore, we require specialised tools, techniques and procedures for securing IoT networks and collecting, preserving and analysing residual evidences of IoT environments. In this special issue we sought new and unpublished works in the domain of IoT security and forensics.

---

* All authors were equally contributed to the paper. Authors' list is ordered alphabetically.

The remainder of this paper is organised as follows. In the next section, we briefly discuss about security challenges in IoT environment. In Section 3, we provide a brief discussion of forensics challenges within IoT networks. We offer a brief review of accepted articles in this special issue in Section 4, and finally conclude the paper.

## 2. Security Challenges in IoT Environments

Wide distribution of IoT nodes and private nature of data that are collected and transferred by IoT devices made security a major challenge. In this section, we are briefly looking at major security challenges that exist in IoT environments.

### 2.1. Authentication

In IoT domain, authentication allows integration of different IoT devices that are deployed in different contexts. Authentication process involves authentication of routing peers that involve in transferring data as well as authentication of the source of data route (data origin node) [7]. Efficient key deployment and key management is a challenge in IoT devices authentication. Any cryptographic key generation and key exchange should not cause a major overhead on IoT nodes [8]. Moreover, in the absence of a guaranteed Certificate Authority (CA), other mechanisms are required for validating cryptographic keys and ensuring integrity of key transfer.

### 2.2 Authorisation and Access Control

Authorisation involves specification of access rights to different resources while Access Control mechanisms should guarantee access right of only authorised resources [9]. Each and every IoT node may only support limited mechanisms for access verification which could be different from other connected objects to the same node [10]. Therefore, deployment and management of a variety of authorisation and access control mechanisms which are tailored to different nodes capabilities is a challenge in a heterogeneous IoT network [11].

### 2.3 Privacy

Deployment of autonomous objects in IoT that sense people private information (such as health data) pose a new level of threat to individuals' privacy. Unlike conventional scenarios in which users have to take some actions (i.e. searching for a keyword or posting some data) to put their privacy at stake, IoT nodes are collecting people's private data without them even noticing [12]. Existing mechanisms are providing user centric privacy, content oriented privacy or context oriented privacy. However, IoT networks are naturally contains autonomous nodes that collect information and require object-oriented privacy models. Moreover, majority of privacy regulations mandate keeping users informed about how their private data is managed and administered. Identifying nodes that may have access to passively collected users' private information is a huge challenge in heterogeneous IoT networks [13].

### 2.4 Secure Architecture

Building an architecture that overcomes aforementioned security challenges in IoT environments is not trivial. Any IoT architecture should not only address previously mentioned security issues but deal with challenges that are introduced by deploying IoT devices over Software Defined Networks (SDN) and cloud infrastructure [14] . Majority of SDN and cloud environment security issues would inevitability inherited to underlying IoT sensors. Moreover, Complexities that involve in securely connecting object-oriented IoT networks to data-oriented cloud infrastructures would introduce many unprecedented security challenges [15]. Finally, detection of malicious traffics rerouted over networks with different natures (i.e. SDN, Cloud and IoT) and hunting for malicious actors is a very challenging task for existing intrusion detection and prevention systems [16].

## 3. Forensics Challenges in IoT Environments

IoT would soon pervade all aspects of our life from managing our home temperature to thinking cars and smart management of the cities. So it won't take long to see people suing each other for misusing their smart things, thinking cars that have accident and attackers who compromised smart sensors. The Internet of everything is developing a haystack which contains lots of valuable forensics artefacts while identification, collection, preservation and reporting of evidences as well as attack or deficit attribution would be challenging in this environment. In this section we briefly introduce main forensics challenges in IoT environments.

### 3.1 Evidence Identification, Collection and Preservation

Search and seizure is an important step in any forensics examination. However, detecting presence of IoT systems is quite a challenge considering these devices are designed to work passively and autonomously [17]! Even, in most cases when an IoT device is identified there is no documented method or a reliable tool to collect residual evidences from the device in a forensically sound manner [18]. Moreover, there are very limited methods to create forensic image of a given IoT device ignoring ethical considerations when collecting evidences from devices running in a multi-tenancy environment.

While preservation of collected data using traditional techniques such as hashing is not difficult, preservation of the scene is an enormous challenge in an IoT environment. Real-time and autonomous interactions between different nodes, would make it very difficult if not impossible to identify scope of a compromise and boundaries of a crime scene.

### 3.2 Evidence Analysis and Correlation

Majority of IoT nodes are not storing any metadata including temporal information which makes provenance of evidences a challenge for an investigator! In the absence of temporal information such as modified, accessed and created time, correlation of evidences gathered from different IoT devices is almost impossible. Beyond technical challenges, privacy is a major issue to consider when analysing and correlating collected data especially as majority of IoT sensors are collecting innate personal information [19]. Moreover, the sheer volume of data that are collected in heterogeneous IoT environments make it close to impossible to provide an end-to-end analysis of residual evidences.

### 3.3 Attack or Deficit Attribution

A common outcome of any forensics investigation is to identify criminal actors or liabilities of involved parties in the case of an incident. With fast rate of development in autonomous vehicles industry identifying liabilities of different parties (i.e. human driver or car autonomous driving system) in an accident would soon become a cyber forensics challenge! Answering such questions would be impossible in the absence of documented methods and forensically sound tools for collection, preservation and analysis of cyber physical systems data [20]. Moreover, in the absence of a proper authentication system, identifying activities and liabilities of different parties having access to an IoT node would be challenging. Finally, attribution of malicious activities detected in an IoT environment even in the possession of evidences is quite challenging in the absence of a reliable and secure architecture that guarantees a forensically sound logging and monitoring system.

## 4. A Brief Review of Accepted Articles of this Special Issue

In this special issue we have accepted three papers in the domains of IoT privacy, IoT forensics and security in SDN-based IoT networks.

The first paper is Privacy-Preserving Protocols for Secure and Reliable Data Aggregation in IoT-Enabled Smart Metering Systems by Samet Tonyali, et al. [21]. This paper tackles privacy issues raised by frequent data collection of smart metering systems. Current systems achieve privacy by concealing aggregated in-network data using Fully Homomorphic Encryption (FHE) and secure MultiParty Computation (secure MPC). However, both FHE and secure MPC are producing overhead in IoT environments. Therefore, Samet Tonyali, et al. [21] suggested a new protocol which utilises FHE and secure MPC in Smart Grid (SG) Advanced Metering Infrastructure (AMI) to reduce overheads while providing a viable privacy-preserving data aggregation mechanism.

The Digital Forensic Intelligence: Data Subsets and Open Source Intelligence (DFINT+OSINT): a Timely and Cohesive Mix paper by Darren Quick and Kim-Kwang Raymond Choo [22] addresses the issue of analysing big digital forensics data resulted from investigation of IoT objects. The paper presents a framework for entity identification and open source information cohesion within digital forensics domain. Authors analysis using real world data demonstrated benefits of their framework in real world investigations.

The last paper of this special issue, has suggested a mechanism to change attack surface in SDN-based IoT networks to increase attackers efforts for a successful exploitation. Internet of Things;

Software-Defined Networking; Attack Graphs;Security Modeling by Mengmeng Ge, et al. [23] utilised SDN to reconfigure IoT network topology as a proactive defense mechanism. Their simulated results showed how their mechanisms increased attackers efforts while maintained the average shortest path length in a SDN-IoT network.

## 5. Conclusion

The fast pace of development and nature of IoT environments bring a variety of security and forensics challenges. In this paper, we briefly presented major security and forensics issues along with potentially promising solutions. Papers included in this special issue offer state of art view of privacy, security and forensics challenges in IoT environments along with innovative solutions that paves the way towards secure and forensically sound deployment of IoT networks.

**Acknowledgements**

We would like to sincerely thank all the authors and reviewers for their efforts toward the success of this special issue. Moreover, we like to appreciate all supports provided from the journal office in managing paper submission and editing papers. Finally, we would like to thank the Editor-in-Chief Prof. Peter Sloot.